# On defect-induced magnetism and bang opening in graphene

A. M. Panich

Department of Physics, Ben-Gurion University of the Negev, P.O.Box 653, Be'er Sheva 84105, Israel

**Abstract**

We discuss several mechanisms of magnetism and band opening in graphene produced by its hydrogenation and fluorination that can be examined experimentally.

Recent discovery of graphene initiated an increasing number of publications (mainly theoretical) on electronic and magnetic properties of this material and its modifications. Some of them are very attractive, while the others seem to be a subject of a criticism. E.g., Yazyev and Helm [1] have recently reported on the eventual magnetism in graphene induced by the chemisorbed hydrogen. Their calculations showed that the hydrogen chemisorption gives rise to the strong Stoner ferromagnetism with a magnetic moment of 1 $\mu_B$ per hydrogen atom at all studied concentrations. As known, the appearance of the Stoner ferromagnetism in the system of collective electrons requires large density of states at the Fermi level $D(E_F)$ and large values of the exchange energy parameter $I$ in order to correspond to the Stoner criterion $ID(E_F)>1$. This is typical of 3d transition metals showing large $D(E_F)$. However, it is not characteristic of graphene, which is a semimetal with vanishing density of states at the Fermi energy. Appearance of a defect-induced impurity band near the Fermi level, calculated in ref. [1], is not a sufficient condition for the Stoner mechanism of ferromagnetism, particularly in a slightly hydrogenated graphene discussed by the authors. In such a system the density of states at the Fermi level caused by the impurity band is low, the Stoner criterion (which, by the way, was not estimated in [1]) is hardly realized, and thus the Stoner or "band" ferromagnetism is unlikely. Therefore it was not observed experimentally.

We note that the formation of the covalent C-H bond under the hydrogen chemisorption in the model of ref. [1] does not create a dangling bond with the localized magnetic moment at the neighboring carbon atom. The authors concluded that this defect is characterized by the slight protrusion of the hydrogenated carbon atom and very small displacement of all other neighbor carbon atoms. It means that this atom still has three C-C bonds and holds nearly $sp^2$ electronic configuration. Calculation by Boukhalov et al. [2] supports an intermediate character of the hybridization between $sp^2$ and $sp^3$ and shows that at the chemisorption of single carbon atom, the hybridization is still rather close to $sp^2$. While one of the electrons forms a bond with hydrogen, the other is unpaired and is actually delocalized over the graphene plane. According to the Fermi statistics, its contribution to the magnetic moment of the electron gas is $\frac{\mu_B^2 B_0}{k_B T_F}$ rather than 1 $\mu_B$ as suggested in ref. [1] (here $B_0$ is the applied magnetic field, $\mu_B$ is the Bohr magneton, and $k_B$ is the Boltzmann constant).

More likely, the magnetism in graphene could be caused by the unpaired electron spins of dangling bonds, vacancies and localized edge states. Such an effect is realized in fluorine-graphite intercalation compounds and fluorinated graphites [3,4], in which the formation of the covalent C-F bonds and transformation of the electronic configuration of carbon from $sp^2$ to $sp^3$ is accompanied by occurrence of the dangling bonds. Moreover, the increase in fluorination yields appearance of finite carbon clusters surrounded by fluorocarbon groups; in such clusters the conductivity is lost. Drop in the conductivity and localization effects in such a system are satisfactory explained in the frameworks of the percolation theory. The aforementioned percolation mechanism seems to be valid also for the hydrogenated graphene in the following way: Under graphene hydrogenation, carbon atom is pulled out of the plane due to formation of the covalent C-H bond and transformation from $sp^2$ to $sp^3$ configuration. When such a bond is formed, the corresponding lattice point in the conductive graphene network is blocked and does not take part in conductivity any longer. The critical number of the unblocked points, at which conductivity still exists, is 70% for a two-dimensional honeycomb lattice [5]. When this threshold of percolation is reached, the infinite conductive carbon sheet is divided into the finite clusters separated from each other by non-conductive regions, thus the conductivity of the graphene sheet is lost. According to the percolation theory, the conductivity of the $(C_xH)_n$ sheet should drop to zero at $x\sim3$, causing a highly conductive graphene plane to undergo a semimetal-insulator transformation. Moving from the local picture of site blocking treated by the percolation theory to the electronic band structure language, such a process leads to the opening of a band gap for graphene and semimetal-insulator or semimetal-semiconductor transformation. Analogously, the fluorination of graphene would cause similar semimetal-semiconductor or semimetal-insulator transformation as the hydrogenation does. In such an approach, the gap is tunable since it depends on the hydrogen or fluorine content. This mechanism can be examined experimentally. The aforementioned approach opens a new opportunity for applicatons of graphene in nanoelectronics.